\documentclass[aip,amsmath,amssymb,reprint]{revtex4-1}

\usepackage{graphicx}
\usepackage{dcolumn}
\usepackage{bm}

\usepackage[utf8]{inputenc}
\usepackage[T1]{fontenc}
\usepackage{etoolbox}

\usepackage{newtxmath}

\usepackage{resizegather}
\usepackage{hyperref}
\usepackage{color}
\usepackage{xcolor}
\usepackage[none]{hyphenat}

\hypersetup{
	colorlinks=true,
	linkcolor=blue,
	filecolor=blue,      
	urlcolor=blue,
	citecolor = blue,breaklinks,backref}

\usepackage[bottom=1.5cm, right=1.5 cm, left=2.0cm, top=2.5cm]{geometry}

\begin{document}

\title{ Non-Hermiticity induced thermal entanglement phase transition}
\author{ {Bikashkali Midya} \\ \it{Indian~Institute of Science Education and Research Berhampur, Odisha 760003, India}}


	\begin{abstract}
			\noindent Theoretical analysis of a prototypical two-qubit effective non-Hermitian system characterized by asymmetric Heisenberg $XY$ interactions in the absence of external magnetic fields demonstrates that maximal bipartite entanglement and quantum phase transitions can be induced exclusively through non-Hermiticity. At thermal equilibrium as $T\rightarrow 0$, the system attains maximal entanglement ${C}=1$ for values of the non-Hermiticity parameter greater than a critical value \mbox{$\gamma>\gamma_c=J\sqrt{1-\delta^2}$}, where $J$ denotes the exchange interaction and \(\delta\) represents the anisotropy of the system; conversely, for \mbox{$\gamma < \gamma_c$}, entanglement is nonmaximal and given by \mbox{${C} = \sqrt{1 - (\gamma/J)^2}$}. The entanglement undergoes a discontinuous  transition to zero precisely at \mbox{$\gamma = \gamma_c$}. This phase transition originates from the closing of the energy gap at a non-Hermiticity-driven ground state degeneracy, which is fundamentally different from an exceptional point. This work suggests the use of singular-value-decomposition generalized density matrix for the computation of entanglement in bi-orthogonal~systems.
	\end{abstract}

		\maketitle

 \noindent {\bf Introduction.--}  { Certain non-Hermitian open quantum systems exhibit exceptional point (EP) spectral singularities, where multiple eigenfrequencies coalesce and their corresponding eigenstates become indistinguishable~\cite{Ashida2021,Rotter2009}. These points were initially investigated in semi-classical systems~\cite{Heiss2012}, leading to phenomena such as EP-induced lasing, non-reciprocal dynamics, and enhanced sensing capabilities (refer to the reviews~\cite{Ganainy2018,Feng2017,Ozdemir2019} and references therein). Subsequently, EPs have been identified as producing novel quantum effects such as universal critical phenomena~\cite{Kawabata2017,Dora2019,Xiao2019}, accelerated relaxation process~\cite{Zhou2023}, chiral state transfer~\cite{Sun2024}, and have been observed in various dissipative quantum systems, including solid-state spins~\cite{Wu2019,Wu2024,Wen2020,Zhang2022}, trapped ions~\cite{Ding2021,Ding2022,Cao2023}, ultracold gases~\cite{Zhao2025,Ren2022,Li2019}, and superconducting and quantum photonic qubits~\cite{Naghiloo2019,Chen2021,Gao2025,Xiao2021}. } In contemporary research, non-Hermitian interactions and EPs are emerging as key resources in the field of quantum information science~\cite{Harrington2022}, and the investigation of entanglement and its dynamics~\cite{Li2023,Khandelwal2024,Selim2025,Longhi2025,Feyisa2025,Han2023,Kumar2022,Zhang2024,Deng2024,Zou2022,Ju2019,Liu2025,Qian2025,Lee2014,Turkeshi2023,Chen2014,Kawabata2023,Lima2024,Lima2025,Rottoli2024,Chang2020,Fang2022,Gal2023,PKumar2022,Arboleda2024,Duc2021,Li2024}. Specifically, investigations have revealed nontrivial quantum advantages such as faster-than-Hermitian entanglement generation~\cite{Li2023}, chiral exchange of Bell states~\cite{Khandelwal2024}, and entanglement filtering within non-Hermitian coupled qubit systems~\cite{Selim2025}.
	
These advancements naturally lead to an important question: {\it Can non-Hermiticity alone induce maximal thermal entanglement \cite{Arnesen2001,Kamta2002,Amico2008} and trigger phase transitions in an otherwise Hermitian qubit system that do not exhibit these features in the absence of external magnetic fields}? Here, we provide an affirmative answer. {By examining a minimal model of two asymmetrically coupled spin qubits, it is demonstrated that maximal entanglement can be attained by tuning the effective non-Hermitian parameter beyond the point where the spectral energy gap closes. 	Notably, this gap-closing point is shown to differ from an EP, as the corresponding states remain distinguishable. This finding contrasts with  prior investigations on analogous Hermitian models~\cite{Arnesen2001,Kamta2002,Gunlycke2001,Wang2001}, which suggested that external magnetic fields are requisite for attaining thermal entanglement transition. Furthermore, this work introduces singular-value-decomposition (SVD) generalized thermal states, which accurately captures the entanglement characteristics of non-Hermitian systems.}

	\paragraph*{\bf Theoretical model.--}
	 We consider an effective non-Hermitian system defined by the Hamiltonian \mbox{$H=H_{XY}+H_{NH}$}, where  the Heisenberg $XY$ Hamiltonian for two-qubit systems having the nearest-neighbor interaction is given by~\cite{Amico2008}
	\begin{equation}
	 H_{\rm XY} = 2J \left[ (1+\delta) S_1^x S_2^x + (1-\delta) S_1^y S_2^y \right].
	 	\end{equation} 
	 	Here, the operators $S_n^{x,y,z} = \sigma_n^{x,y,z}/2$, defined in terms of the Pauli matrices, correspond to the local spin-$\tfrac{1}{2}$ operator at qubit~$n$, $J>0$ is the antiferromagnetic exchange interaction, and  the dimensionless parameter $\delta$ $(0 \leq \delta \leq 1)$ denotes anisotropy of the system; specifically, $\delta = 0$ characterizes an isotropic interaction, whereas $\delta = 1$ corresponds to a fully anisotropic Heisenberg-Ising interaction. When $\delta$ equals $1$, the Hamiltonian $H_{XY}$ does not exhibit thermal entanglement \cite{Kamta2002}. Additionally, a quantum phase transition is not observed for values of $\delta<1$ unless an external tunable parameter, such as an applied magnetic field, is introduced \cite{Arnesen2001,Kamta2002,Gunlycke2001,Wang2001}.  The non-Hermitian component of the Hamiltonian is given by 
	 	\begin{equation}
	 		H_{\rm NH} = \gamma_1 S_1^- S_2^+ + \gamma_2 S_1^+ S_2^-,
	 		\end{equation}
	 		 where  \mbox{$S_n^\pm=S_n^x\pm i S_n^y$} are creation and annihilation operators, and $\gamma_1\ne\gamma_2^*$ parameterize the rate of asymmetric exchange between two qubits in the subspace $\{\lvert\uparrow\downarrow\rangle,\lvert\downarrow\uparrow\rangle\}$ spanned by single spin excitations. For analytical tractability, we impose an anti-Hermitian condition \mbox{$H_{NH}^\dag=-H_{NH}$}  satisfied by \mbox{$\gamma_1 = -\gamma_2 = \gamma$}. Under this assumption,  the total Hamiltonian simplifies to
	\begin{equation}
	 H=(J+\gamma)S_1^-S_2^+ +(J-\gamma)S_1^+S_2^-  + J\delta(S_1^- S_2^-+S_1^+ S_2^+). \label{eq-Hamiltonian}
	\end{equation}
	{The Hamiltonian \eqref{eq-Hamiltonian}, apart from an overall dissipative term \mbox{$(-i\gamma)$}, represents the effective Hamiltonian of a fully postselected Markovian open quantum system~\cite{Reiter2012,Minganti2019,Minganti2020}. This Hamiltonian can be derived from the Lindblad master equation (with $\hbar=1$) \cite{Gardiner2004} given by:
		\begin{gather}
			\frac{d\rho(t)}{dt}=\mathcal{L}\rho(t)=-i[H_{XY},\rho]+\mathcal{D}[L_{12}] \rho+\mathcal{D}[G_{12}]\rho. \label{Eq-Liouvillian}
		\end{gather}		
		Here, $\rho$ is the density operator of the system, and $\mathcal{L}$ denotes a hybrid Liouvillian that includes the coupled spin Hamiltonian \(H_{XY}\), which governs the coherent evolution of the system. The operators \(L_{12}\) and \(G_{12}\) are pairwise jump operators~\cite{Takemori2025,Metelmann2015,Song2019} defined as:
		\begin{equation}
			L_{12}=\sqrt{\frac{\gamma}{2}}(S_1^- -i S_2^-), \quad G_{12}=\sqrt{\frac{\gamma}{2}}(S_1^+ +i S_2^+), \label{Eq-Lindblad}
		\end{equation}
		These operators characterize the system’s nonunitary interaction with its environment through a generalized  dissipator~\cite{Minganti2020}:
		\begin{equation}
			\mathcal{D}[Z]\rho= 2 q Z\rho Z^{\dagger} - (Z^\dagger Z \rho+\rho Z^\dagger Z).
		\end{equation}
	Here, $\gamma>0$ represents the rate of dissipative interaction, while the parameter \(q \in [0,1]\) controls the degree of postselection dynamics: \(q=0\) corresponds to full postselection dynamics, whereas \(q=1\) indicates no postselection. The Liouvillian in equation \eqref{Eq-Liouvillian} can be expressed as
	\begin{equation}\label{Eq-Liouvillian-1}
		\mathcal{L}\rho = -i(H_{\rm{eff}}\rho - \rho H_{\rm{eff}}^\dag) + q(2L_{12}\rho L_{12}^\dag + 2G_{12}\rho G_{12}^\dag),
	\end{equation}
	where the effective non-Hermitian Hamiltonian reduces to
	\begin{equation}\label{Eq-Heff}
		H_{\rm eff} = H_{XY} - i L_{12}^\dag L_{12} - i G_{12}^\dag G_{12} = H - i \gamma.
	\end{equation}
	This final equality is derived using the relations \(\{S_1^-, S_1^+\} = \{S_2^-, S_2^+\} = 1\) and \([S_1^-, S_2^+] = 0\). When postselected trajectories of null quantum jump is chosen (i.e., \(q=0\)), equations \eqref{Eq-Liouvillian-1} and \eqref{Eq-Liouvillian} show that the Lindblad master equation simplifies to the von Neumann equation. In this scenario, the dynamical solution given by \(\rho(t) = e^{-itH_{\rm eff}} \rho(0) e^{itH_{\rm eff}^\dag}\) is formally equivalent to a thermal state \(\rho(T)\) through the Wick rotation \(it = \frac{1}{k_B T}\).  Note that the uniform background loss term \((-i\gamma)\) in equation \eqref{Eq-Heff} does not affect the thermal entanglement discussed later; therefore, it has been omitted, leading to the approximation \(H \simeq H_{\rm eff}\).
	}
				
It may be noted that postselected non-Hermitian quantum systems have been experimentally realized in superconducting qubits \cite{Naghiloo2019,Chen2021}. The effective Hamiltonian described in Eq.~\eqref{eq-Hamiltonian} is also significant in cascaded qubit networks, such as when  qubits are coupled to a chiral bath~\cite{Pichler2015,Stannigel2012,Metelmann2015}.

\begin{figure}[t!]
	\centering{\includegraphics[width=0.48\textwidth]{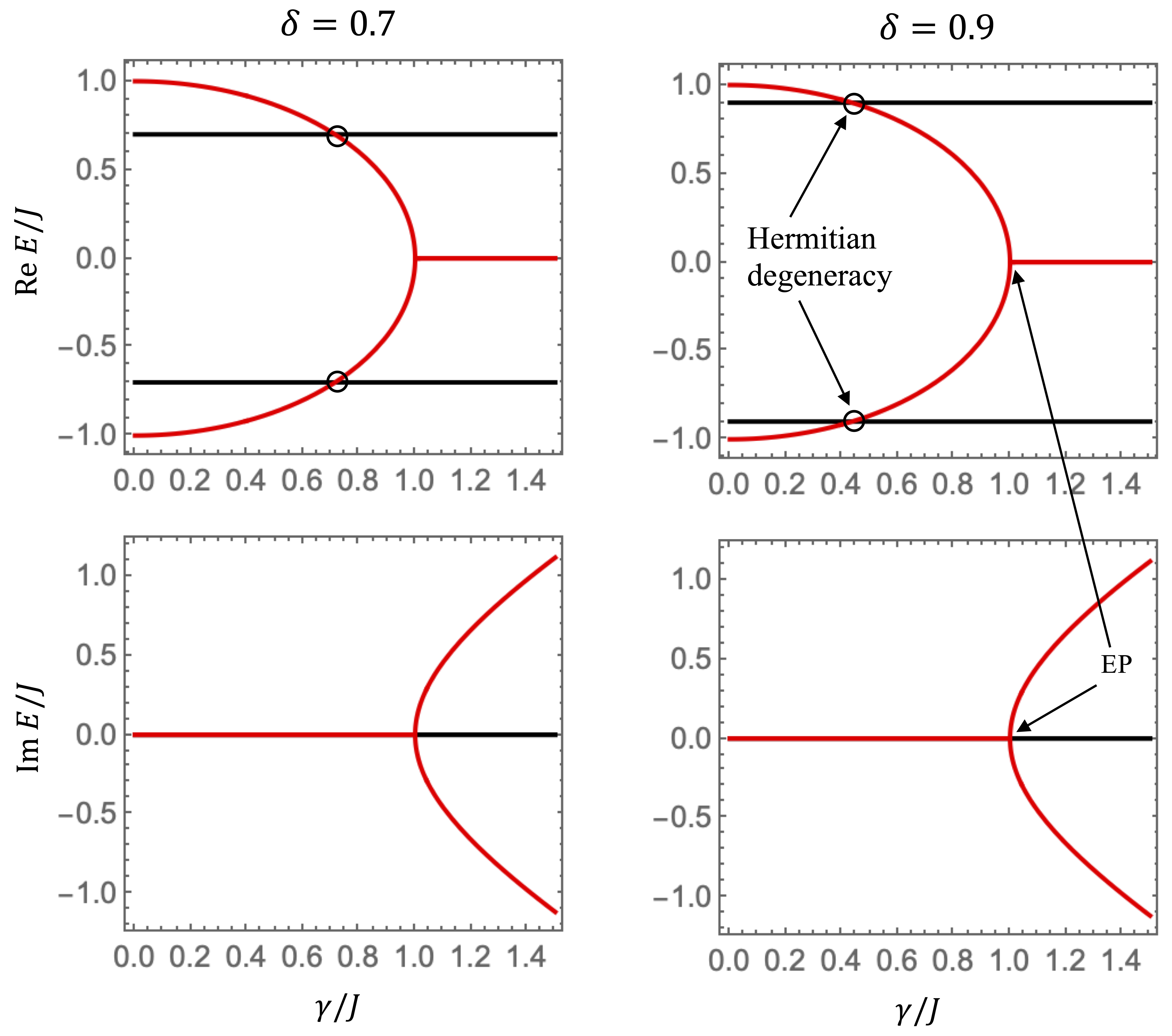}}
	\caption{{\it Non-Hermiticity induced Hermitian degeneracy.} Energy spectra of \(H\) for two distinct values of \(\delta\) illustrate the appearance of a Hermitian degeneracy induced by non-Hermiticity, which differs from an exceptional point (EP). When the parameter \(\gamma\) is varied, the ground state and the first excited state interchange their positions. It is also observed that ground-state-interchange takes place at smaller values of $\gamma$ when $\delta$ is larger. This state switching plays a crucial role in the thermal entanglement transition described in the main text. Here, $J=1$ is chosen. } \label{fig-s1}
\end{figure}

	\paragraph*{\bf Energy spectrum and degeneracy.--} The  right- and left-eigenstates of the Hamiltonian $H$ are expressed in the standard two-qubit basis $\{\lvert\uparrow\uparrow\rangle,\lvert\uparrow\downarrow\rangle,\lvert\downarrow\uparrow\rangle,\lvert\downarrow\downarrow\rangle\}$:
	\begin{align}
		& 	\scalebox{.95} {$ 	|R_{0,3}\rangle: \frac{1}{\sqrt{2}}\left( \sqrt{\frac{J-\gamma}{J+\gamma}}\lvert\uparrow\downarrow\rangle \mp \lvert\downarrow\uparrow\rangle \right),	|R_{1,2}\rangle: \frac{1}{\sqrt{2}}\left( \lvert\uparrow\uparrow\rangle \mp \lvert\downarrow\downarrow\rangle \right) $} \nonumber\\ 
		&  \scalebox{.95} {$ 	|L_{0,3}\rangle: \frac{1}{\sqrt{2}}\left( \sqrt{\frac{J+\gamma}{J-\gamma}}\lvert\uparrow\uparrow\rangle \mp \lvert \downarrow\downarrow\rangle \right),  ~ \lvert L_{1,2}\rangle = \lvert R_{1,2}\rangle $},
	\end{align}
	which satisfy the bi-orthonormality condition \cite{Brody2013} \mbox{$\langle L_j|R_{j'}\rangle=\delta_{jj'}$}, and fulfill completeness relation \mbox{$\sum_j |R_j\rangle\langle L_j|=I$} away from an exceptional point $(\gamma=J)$. These eigenstates correspond to a purely real energy spectrum across all values of $\delta$:
	\begin{equation}
	E_{0,1,2,3}=\{-\sqrt{J^2-\gamma^2},-J \delta,J \delta, \sqrt{J^2-\gamma^2}\}, \label{eq-specta}
	\end{equation}
	 provided the non-Hermitian parameter satisfies the inequality \mbox{$\gamma< J$}. It may be noted that the non-Hermiticity affects only the states \(|R_{0}\rangle\) and \(|R_{3}\rangle\) corresponding to the energy levels \(E_0\) and \(E_3\), respectively. As the parameter \(\gamma\) increases, these two energy levels move closer together (see Fig.~\ref{fig-s1}). At \(\gamma = J\), the system reaches an exceptional point (EP) where both energies \(E_0\) and \(E_3\) and their associated eigenstates \(|R_{0}\rangle\) and \(|R_{3}\rangle\) merge. For values of \(\gamma > J\), a pair of complex conjugate energy levels emerges. In addition to this known EP degeneracy, a novel degeneracy occurs within the intermediate range \(0 \leq \gamma < J\), where all energies remain real. This new degeneracy is characterized by two distinct real energy levels becoming equal while their corresponding eigenstates stay distinct and orthogonal. Specifically, the energy gaps between the pairs \((E_0, E_1)\) and \((E_2, E_3)\) simultaneously close when the condition \(\gamma / J = \sqrt{1 - \delta^2}\) holds. Figure~\eqref{fig-s1} presents the full energy spectrum of \(H\) for different anisotropy parameters \(\delta\), demonstrating how this Hermitian degeneracy arises as the non-Hermiticity parameter \(\gamma\) is varied. This phenomenon, termed ‘non-Hermiticity assisted Hermitian degeneracy’ and distinct from an EP, plays a crucial role in controlling low-temperature thermal entanglement and phase transitions, as explained below. In this paper, we do not discuss entanglement in the situation of complex spectrum \mbox{$E^R_j=(E^{L}_j)^*$} for \mbox{$\gamma>J$}, in order to preclude non-unitary dynamical evolution.

	\paragraph*{ \bf Non-Hermiticity induced thermal entanglement and phase transition.--}	To examine the thermal entanglement in the system, we employ the bi-orthogonal density operator in thermal equilibrium defined as \cite{Arnesen2001,Brody2013} \mbox{$\rho(T)={Z}^{-1} e^{-H/k_BT}$}, where 
	\begin{equation}
		e^{-H/k_BT}=\sum_{j=0}^3 e^{-E_j/k_BT} |R_j\rangle\langle L_j|,
	\end{equation}
and	\mbox{${Z}={\rm Tr}~e^{-H/k_B T}$} denotes the partition function. We have used the above-mentioned biorthogonal completeness relation and the trace is defined as \mbox{${\rm Tr} (\cdot)=\sum_j\langle L_j\rvert \cdot\lvert R_j\rangle$}. Here, $T$ is the temperature and $k_B$ is the Boltzmann constant (which is set to unity). The density matrix is explicitly given by
	\begin{gather}\label{eq-rho}
		\rho = \frac{1}{{Z}} \left[\begin{array}{cccc}
			\cosh \frac{J\delta}{T}  &  0&0  & -\sinh \frac{J\delta}{T} \\
			0& \cosh \frac{  \sqrt{J^2-\gamma^2}}{T}&-\sqrt{\frac{J-\gamma}{J+\gamma}}\sinh \frac{  \sqrt{J^2-\gamma^2}}{T}& 0 \\
			0& -\sqrt{\frac{J+\gamma}{J-\gamma}}\sinh \frac{  \sqrt{J^2-\gamma^2}}{T} &\cosh \frac{  \sqrt{J^2-\gamma^2}}{T}&0  \\
			-\sinh \frac{J\delta}{T} & 0 & 0 & \cosh \frac{J\delta}{T}
		\end{array} \right],
	\end{gather}
	where \mbox{${Z}=2(\cosh \frac{\delta J}{T} + \cosh \frac{\sqrt{J^2-\gamma^2}}{T})$}, is non-Hermitian \mbox{$\rho^\dag\ne\rho$}. To ensure consistent computation of entanglement measure ({see the detailed discussion in Appendix~\ref{Appendix-A} and the accompanying figure~\eqref{fig-s2}}) , here we introduce the SVD generalized density matrix \cite{Parzygnat2023}
	\begin{gather}
		\rho^{SVD}(T)= \frac{\sqrt{\rho^\dag(T)\rho(T)}}{{\rm Tr}~\sqrt{\rho^\dag(T)\rho(T)}} =\frac{\sqrt{e^{-H^\dag/k_BT}e^{-H/k_BT}}}{{\rm Tr}~\sqrt{e^{-H^\dag/k_BT}e^{-H/k_BT}}}.
	\end{gather}
	Note that \mbox{$\rho^{SVD}=\rho$} when $H$ is Hermitian. The degree of entanglement between two qubits is quantified by the concurrence \cite{Wooters1998} defined by \mbox{$C=\max\{\lambda_0-\lambda_1-\lambda_2-\lambda_3,0\}$}, where $\lambda_j$ are non-negative eigenvalues, arranged in decreasing order, of the operator 
	\begin{gather}
		R=\left[\rho^{SVD}(\sigma^y\otimes\sigma^y){\rho^{SVD}}^*(\sigma^y\otimes\sigma^y)\right]^{\frac{1}{2}}.
	\end{gather}
	The concurrence ranges from $0$ to $1$, with a value of zero indicating the absence of entanglement and a value of one corresponding to maximal entanglement between two qubits. As shown in Appendix-\ref{Appendix-A}, for (non-degenerate) pure states \mbox{$\rho_j=\lvert R_j\rangle\langle L_j\rvert$}, the concurrence  
	\begin{equation}
C(\rho_j)=\sqrt{1-(\gamma/J)^2}, \quad  j=0,3,
	\end{equation}
	 and \mbox{${C}(\rho_j)=1$} for \mbox{$j=1,2$}.  This indicates that, while the energy eigenstates $\lvert R_{1}\rangle$ and $|R_2\rangle$ are maximally entangled Bell states, the states $|R_{0}\rangle$ and $|R_{3}\rangle$ exhibit non-maximal entanglement in the presence of non-Hermiticity; in fact, they become separable at the exceptional point $\gamma=J$. 
	
		\begin{figure*}[t]
		\centering
	\includegraphics[width=0.62\textwidth]{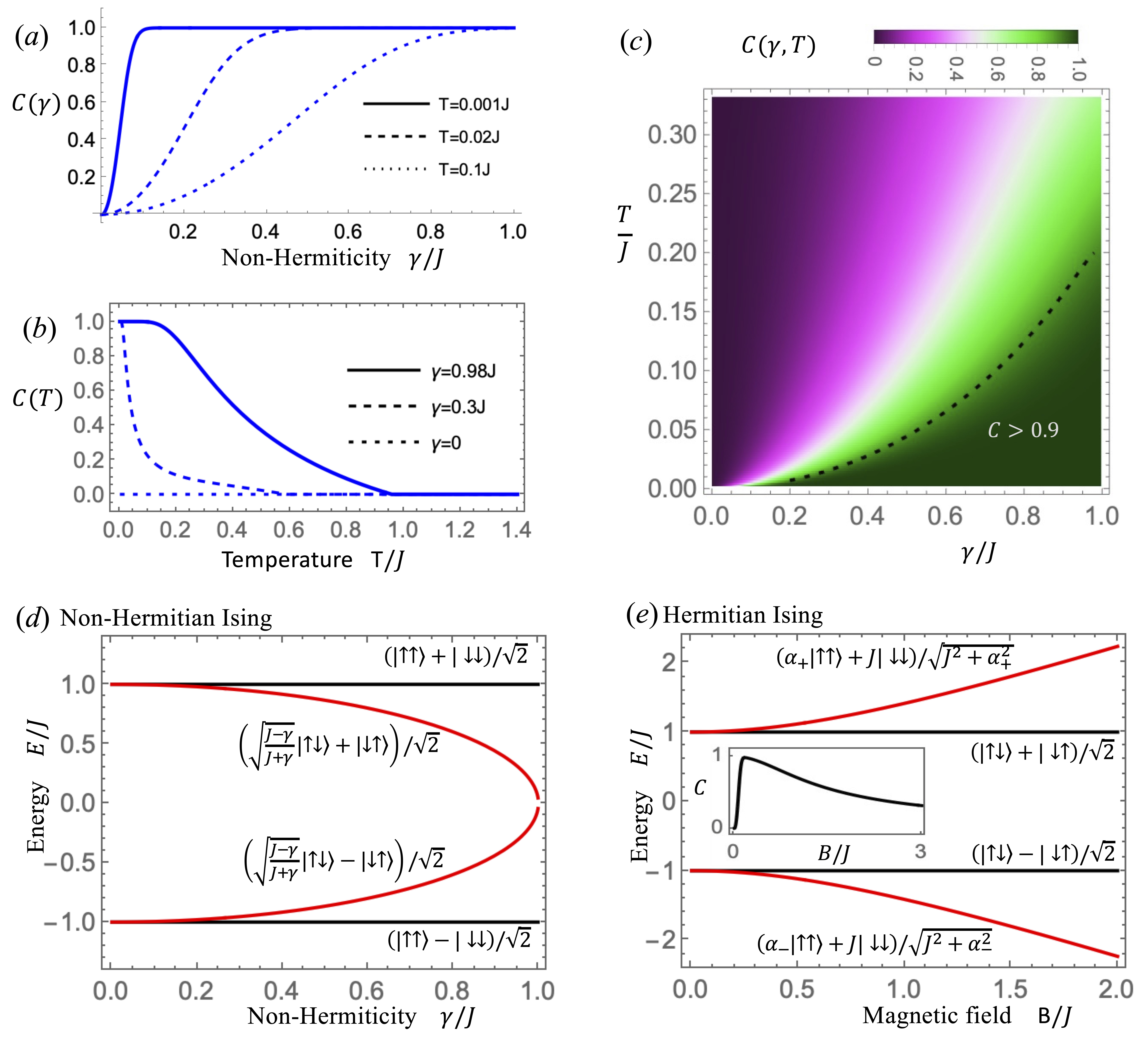}
		\caption{{\it Thermal entanglement in the non-Hermitian Heisenberg-Ising model $(\delta=1)$.}  (a) The concurrence $C$, calculated from the thermal mixed state $\rho$ as defined in Eq.~\eqref{eq-rho}, is presented as a function of $\gamma$ at three distinct temperatures. At temperature near zero, entanglement reaches its maximum for non-zero values of $\gamma$, whereas at finite temperatures, stronger non-Hermiticity is required to achieve comparable entanglement. In panel (b), the concurrence ${C}$ is plotted against temperature for three different values of $\gamma$, illustrating an exponential decay of entanglement with increasing temperature. Panel (c) depicts ${C}$ within the entire parameter range $0 \leq T \leq J/3$ and $0 \leq \gamma < J$. The region below the dashed line represents concurrence ${C}>0.9$, follows from  Eq.~\eqref{eq-7}.  Maximal entanglement observed at $T=0$ originates from the non-Hermiticity-assisted non-degenerate ground state, which corresponds to a Bell state, as shown in panel (d). This behavior contrasts sharply with that of the Hermitian system subjected to an external transverse field $B \hat{z}$, described by \mbox{$H = H_{XY} + B(S_1^z + S_2^z)$}; its  ground state is non-maximally entangled. The corresponding energy spectrum, \mbox{$\{\pm\sqrt{J^2+B^2},\pm J\}$}, and zero temperature concurrence are provided in panel (e) for comparison, where \mbox{$\alpha_{\pm} = B \pm \sqrt{J^2 + B^2}$}. }\label{fig-1}
	\end{figure*}

\begin{figure*}[t!]
	\centering{\includegraphics[width=0.65\textwidth]{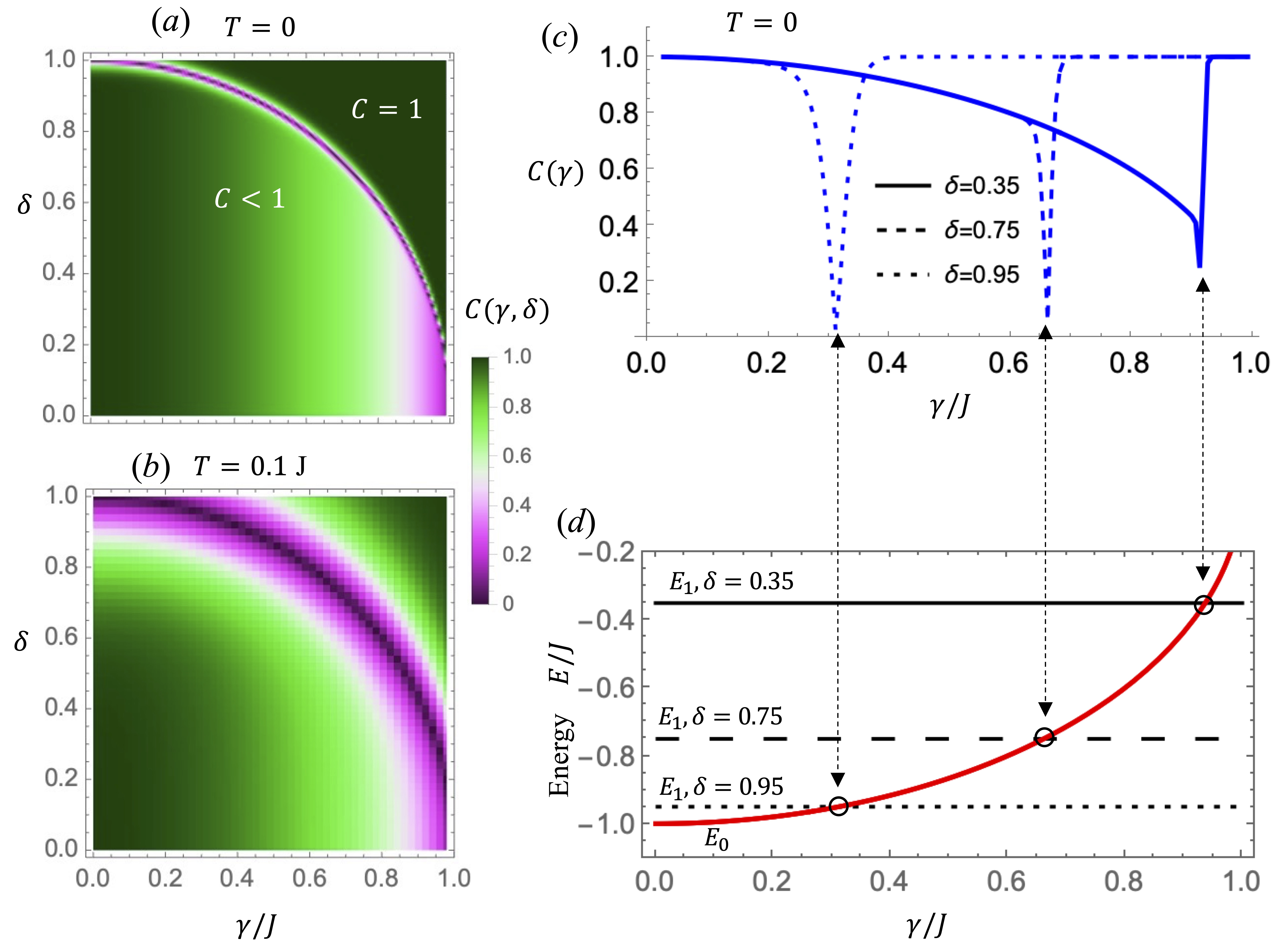}}
		\caption{{\it Non-Hermiticity induced thermal entanglement phase transition.}   	Thermal concurrence at absolute zero temperature ($T=0$) and at $T=0.1J$ is shown in panels (a) and (b), respectively, across the full range of the anisotropy parameter $0 \leq \delta \leq1$ and non-Hermiticity $0 \leq \gamma < J$. The results indicate that entanglement experiences a discontinuous transition from $C=1$ when $\gamma > \gamma_c$ to $C=0$ exactly at the critical point \mbox{$\gamma_c = J(1-\delta^2)^{1/2}$}, followed by an asymptotic increase for values of $\gamma < \gamma_c$. This sharp discontinuity in zero temperature entanglement at the critical non-Hermiticity $\gamma_c$ serves as a hallmark of  quantum phase transition. Panel (c) shows that the discontinuity shifts toward lower values of $\gamma$ as $\delta$ increases. Panel (d) shows that the entanglement phase transition coincides with the occurrence of non-Hermiticity-assisted Hermitian degeneracy where $E_0 = E_1$ (marked by circles), where a change in the ground state from a non-maximally entangled state to a maximally entangled state as a function of $\gamma$ occurs. }\label{fig-2}
	\end{figure*}

	The entanglement characteristics in the thermally mixed state described by equation \eqref{eq-rho} exhibit  richer complexity. In general, the concurrence valid for all temperature and system parameters cannot be studied analytically.  Numerically computed results obtained from Eq.~\eqref{eq-rho} are exemplified in Fig.~\ref{fig-1} and Fig.~\ref{fig-2}. To gain analytical insight into the system's low-temperature entanglement, here, we approximate the mixed state by considering only lowest populated ground and first excited states with Boltzmann weight \mbox{$e^{-E_0/T}/(e^{-E_0/T}+e^{-E_1/T})$} and \mbox{$e^{-E_1/T}/(e^{-E_0/T}+e^{-E_1/T})$}, respectively. This approximation is valid within the temperature range \mbox{$0\le T\lesssim |E_1-E_0|$} and away from the exceptional point (i.e. at $\gamma=J$, where the second excited state also becomes relevant). In this case, eigenvalues $\lambda_{0,1,2,3}$ of $R$ are given by  (detailed calculations are provided in Appendix-\ref{Appendix-B})
	\begin{equation}\label{eq-eigv}
		 \left\{\lambda e^{J\delta/T}, \lambda e^{\sqrt{J^2-\gamma^2}/T}, 0, 0\right\},
	\end{equation}
	when \mbox{$J\delta>\sqrt{J^2-\gamma^2}$},  whereas the first two eigenvalues switch their positions if $J\delta<\sqrt{J^2-\gamma^2}$. Here, \mbox{$\lambda=\sqrt{J^2-\gamma^2}/(Je^{\sqrt{J^2-\gamma^2}/T}+\sqrt{J^2-\gamma^2}e^{J\delta/T})$}. The concurrence \mbox{${C}=\max \{\lambda_0-\lambda_1,0\}$} reduces to 
	\begin{gather}\begin{array}{ll}
			{C}(T)&=\frac{\sqrt{J^2-\gamma^2}}{J e^{\frac{\sqrt{J^2-\gamma^2}}{T}}+\sqrt{J^2-\gamma^2} e^{\frac{J\delta}{T}}} |e^{\frac{\sqrt{J^2-\gamma^2}}{T}}-e^{\frac{J\delta}{T}}|\\ \\
			&	\underset{T\rightarrow0}{=} \left\{\begin{array}{c}
				\sqrt{J^2-\gamma^2}/J, \hspace{0.225cm} \gamma<J\sqrt{1-\delta^2}	\\
				0, \hspace{2cm} \gamma=J\sqrt{1-\delta^2}	\\
				1, \hspace{2cm} \gamma>J\sqrt{1-\delta^2}
			\end{array} \right.
		\end{array}\label{eq-QPT}
	\end{gather}
	Equation \eqref{eq-QPT} represents a key finding, and is valid for all values of $\delta$, \mbox{$0\le\gamma<J$} and $T\sim0$. The  following important conclusions are drawn from this equation.
	
	$i$. First, we discuss entanglement of thermal states in the Heisenberg-Ising system \mbox{$(\delta=1)$}. Interestingly, the concurrence $C$ reaches 1 at absolute zero temperature ($T=0$) whenever $\gamma \gtrsim 0$.  The emergence of maximal entanglement at  zero temperature in the non-Hermitian Ising model can intuitively be explained  by analyzing the ground state characteristics. For $\gamma=0$, corresponding to a Hermitian system, the thermal state is separable and consists of an equal mixture of degenerate ground and first excited eigenstates, both of which are maximally entangled \cite{Gunlycke2001,Kamta2002}. The introduction of non-Hermiticity lifts this degeneracy in a manner distinct from the effect caused by an external magnetic field in Hermitian systems, as illustrated in Figures \ref{fig-1}d and \ref{fig-1}e.  Specifically, increasing $\gamma$ causes the first and second excited states within the single spin excitation subspace \mbox{$\{\lvert\uparrow\downarrow\rangle, \lvert\downarrow\uparrow\rangle\}$} to approach an exceptional point. This transition elevates the maximally entangled triplet Bell state $\left( \lvert\uparrow\uparrow\rangle - \lvert\downarrow\downarrow\rangle\right)/\sqrt{2}$ to become the ground state of the system. This effect is termed non-Hermiticity-induced maximal thermal entanglement and represents a mechanism fundamentally different from magnetically induced entanglement found in Hermitian Ising models (see, for example, Ref.~\cite{Gunlycke2001}).

	$ii$.  For \(\delta=1\) and a specified $\gamma$, the first equation of Eq.~\eqref{eq-QPT} indicates that the system generates concurrence~\mbox{$> {C}$}, for all temperatures bounded by 
	\begin{gather}\label{eq-7}
		T< (J^2-\sqrt{J^2-\gamma^2})/\ln\frac{{C}J+\sqrt{J^2-\gamma^2}}{(1-{C})\sqrt{J^2-\gamma^2}}.
	\end{gather}
	The above inequality provides the low-temperature estimation  of $\gamma$ and $T$ for a desirable ${C}$. Numerical results shown in Figs.~\ref{fig-1}a and \ref{fig-1}c demonstrate that achieving  comparable entanglement at finite temperatures requires stronger non-Hermiticity.

	$iii$. In systems exhibiting anisotropy with \mbox{$0<\delta<1$}, the presence of non-Hermiticity leads to a quantum phase transition marked by an abrupt behaviour change from non-maximal to maximal entanglement at zero temperature ($T=0$). This transition takes place exactly at the critical point \mbox{$\gamma_c/J = \sqrt{1 - \delta^2}$}, as shown in Figures~\ref{fig-2}a and \ref{fig-2}c. The underlying mechanism can be understood by examining the ground state characteristics: for values of $\gamma$ below $\gamma_c$, the ground state is a non-maximally entangled singlet state $|R_0\rangle$. When $\gamma$ surpasses $\gamma_c$, the ground state shifts to $|R_1\rangle$, which is a maximally entangled triplet Bell state. At the critical threshold $\gamma=\gamma_c$, where entanglement drops to zero, this coincides with the spectral degeneracy condition $E_0 = E_1$ (refer to Figures~\ref{fig-2}c and \ref{fig-2}d). Furthermore, an increase in the anisotropy parameter $\delta$ corresponds to a decrease in the critical value of $\gamma_c$ at which this phase transition occurs. Numerical results presented in figure~\ref{fig-2}b also indicate that this behavior of entanglement phase transition remains observable at finite temperatures.

	$iv$. In the case of isotropic Heisenberg interaction ($\delta=0$), the entanglement decreases as $\gamma$ increases. When $\gamma$ reaches the value $J$, all four energy levels become degenerate, resulting in the system being maximally mixed with the density matrix given by $\rho=\mathbb{I}/4$ (see Eq.~\eqref{eq-rho}).

{In general, for arbitrary values of \(\gamma_1\) and \(\gamma_2\), the ground state of \(H\) is the Bell state $\lvert R_1\rangle$ when the condition $E_1<E_0$ holds, where \mbox{$E_0=-\sqrt{(J+\gamma_1)(J+\gamma_2)}$} and \mbox{$E_1 =- J \delta$}.  If this condition is not met, the ground state corresponds to a nonmaximally entangled state characterized by its concurrence 
	\begin{equation}
		C = \frac{2\sqrt{(J+\gamma_1)(J+\gamma_2)}}{2J+\gamma_1+\gamma_2}. \label{Eq-Concurrence}
\end{equation}
 Therefore, the entanglement behavior at zero temperature changes along the energy-gap-closing contours defined by $E_0=E_1$.  A phase diagram illustrating this behavior of entanglement transition for various values of the parameters $(\gamma_1,\gamma_2,\delta)$  is presented in figure~\eqref{fig-3}, whereas complete eigensolutions of \(H\) are discussed in Appendix \ref{Appendix-C}. 
}

The ground-state entanglement transition can also be induced by considering the following effective non-Hermitian Hamiltonian
	\begin{equation}\label{Eq-Hamiltonian-1}
		\bar{H}=H_{XY}-i\gamma S_1^+S_1^- -i\gamma S_2^-S_2^+,
	\end{equation}
		 which correspond to an open system with Lindblad jump operators \mbox{$L_1=\sqrt{\gamma}S_1^- $} and \mbox{$L_2=\sqrt{\gamma}S_2^+$} acting locally on qubit 1 and qubit 2, respectively. The real part of the spectrum of \(\bar{H}\) can be easily confirmed to be isospectral with that of \(H\) in Eq.~\eqref{eq-specta}:~\mbox{$\text{Spec}(\bar{H})=\text{Spec}(H)-i\gamma$}. Also, the entanglement properties of both \(H\) and \(\bar{H}\) are also comparable. Specifically, the ground state spectral and entanglement transition occurs at \(\gamma/J = \sqrt{1 - \delta^2}\). Beyond this transition point, the ground state becomes independent of the non-Hermitian parameter \(\gamma\) and corresponds to the Bell state \(\lvert R_1\rangle\).  Thermal entanglement in models similar to Eq.~\eqref{Eq-Hamiltonian-1}  has been examined in the presence of external fields in references \cite{YueLi2023,Yunpeng2025}. 
		 
		  \begin{figure}[t!] 
			\centering{\includegraphics[width=0.36\textwidth]{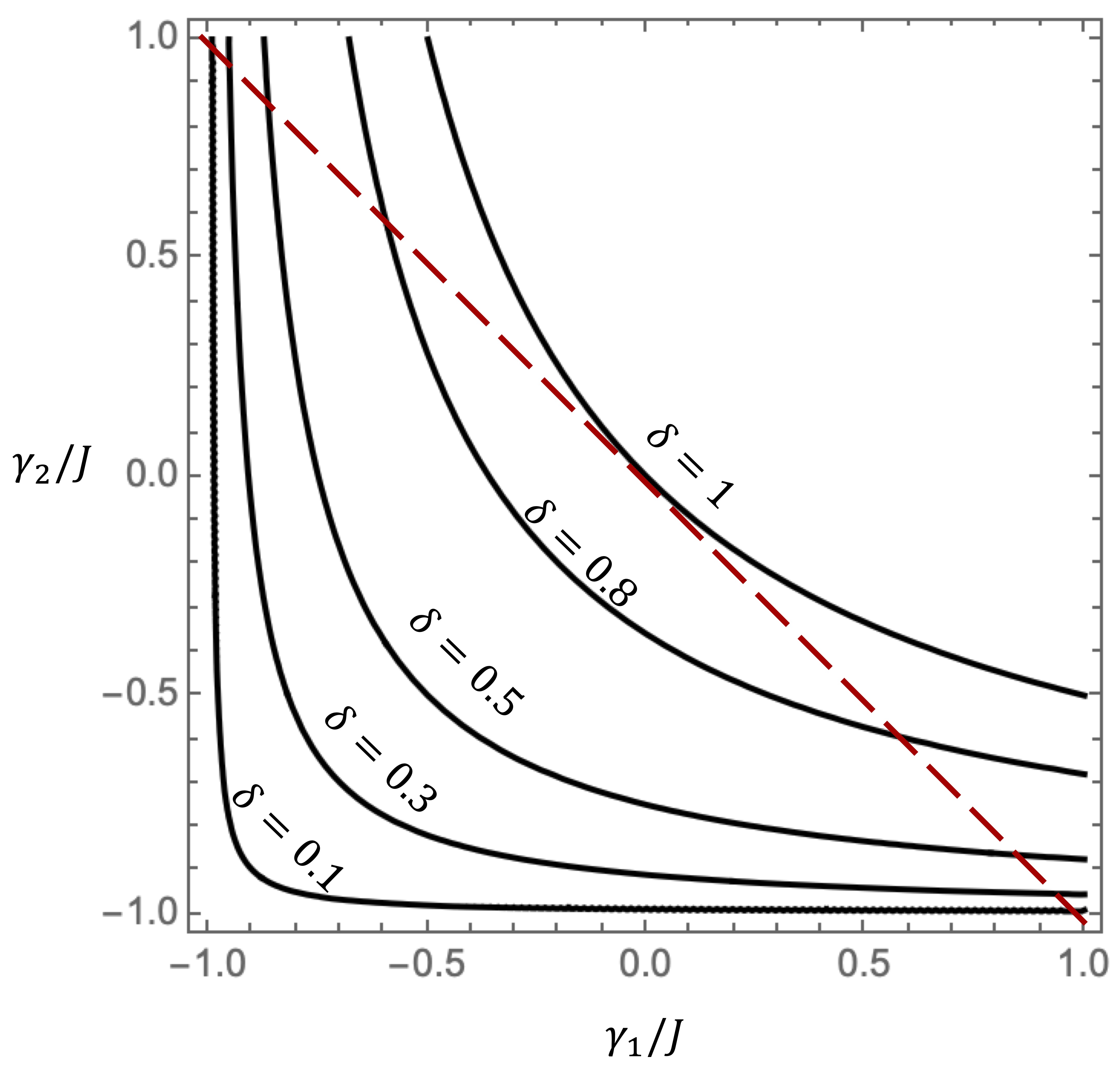}}
			{\caption{The contour lines defined by the equation \mbox{$(J+\gamma_1)(J+\gamma_2) = J^2 \delta^2$} in the $(\gamma_1, \gamma_2)$ parameter plane indicate the transition in ground-state entanglement from maximal to nonmaximal values for different anisotropy \(\delta\). For each specific \(\delta\), the entanglement remains maximal below the corresponding contour line, while above it, the entanglement becomes nonmaximal. The special case of \mbox{$\gamma_1=-\gamma_2$} is shown by the dashed line. } \label{fig-3}}
		\end{figure}
	
	\paragraph*{\bf Conclusion.--} Although the experimental realization of effective non-Hermitian spin models, such as that described by Eq.~\eqref{eq-Hamiltonian}, remains uncertain, theoretical analysis reveals that non-Hermitian interactions alone are capable of inducing maximal bipartite thermal entanglement and its associated phase transition in the absence of external magnetic fields. {The method of postselected trajectories of no quantum jumps is considered essential for observing these phenomena in the steady-state dynamics. This necessity arises because the ground states of effective Hamiltonians studied here do not correspond to the dark states of the associated Liouvillian. As a result, stochastic interactions with the environment significantly alter the entanglement characteristics of the system compared to those predicted by an effective Hamiltonian. Extending this theoretical framework to postselected systems comprising a larger number of spins and exploring many-body entanglement~\cite{Amico2008,Abanin2019} merit further research.}


		\appendix

		\section{\label{Appendix-A} Why $\rho^{SVD}?$ }
		Here we elaborate with a simple example why $\rho^{\rm SVD}$ is suitable in the computation of entanglement  in a bi-orthogonal system.  Consider a bi-orthogonal eigenstate, such as $|R_{0}\rangle= \frac{1}{\sqrt{2}}\left( \sqrt{\frac{J-\gamma}{J+\gamma}}\lvert\uparrow\downarrow\rangle - \lvert\downarrow\uparrow\rangle \right)$. It is evident that this state becomes separable at the EP $\gamma=J$;  otherwise, it remains non-separable and thus entangled. Additionally, the degree of entanglement is expected to decrease as $\gamma$ increases. To quantify the entanglement, we first consider two-qubit bi-orthogonal density matrix:
		\begin{equation}\begin{array}{ll}
			\rho &= |R_0\rangle \langle L_0| \\
			&=\frac{1}{2}\left( \lvert\uparrow\downarrow\rangle\langle\uparrow\downarrow\rvert-\sqrt{\frac{J-\gamma}{J+\gamma}} \lvert\uparrow\downarrow\rangle\langle\downarrow\uparrow\rvert- \sqrt{\frac{J+\gamma}{J-\gamma}} \lvert\downarrow\uparrow\rangle\langle\uparrow\downarrow\rvert+ \lvert\downarrow\uparrow\rangle\langle\downarrow\uparrow\rvert\right),
			\end{array}
		\end{equation}
	which is in a matrix form given by 
		\begin{gather}
			\rho= \frac{1}{2}\left(\begin{array}{cccc}
				0 &  0&0  & 0 \\
				0&1 &- \sqrt{\frac{J-\gamma}{J+\gamma}}& 0 \\
				0&  - \sqrt{\frac{J+\gamma}{J-\gamma}}& 1 &0  \\
				0& 0 & 0 & 0
			\end{array} \right),\label{Eq-B1}
		\end{gather}
		which is non-Hermitian \mbox{$\rho^\dag\ne\rho$}.  The reduced density matrix  \mbox{$\rho^1=\rm Tr_2(\rho) =\frac{1}{2}\left(\lvert\uparrow\rangle\langle\uparrow\rvert+\lvert\downarrow\rangle\langle\downarrow\rvert\right)$}, for the first qubit, corresponds to the  von-Neumann entropy \mbox{$S=-x_1\log_2 x_1-x_2\log_2 x_2=1$}, where $x_1=x_2=\frac{1}{2}$ are the eigenvalues of $\rho^1$.  This shows that the system is maximally entangled irrespective of the strength of non-Hermiticity defined by $\gamma$. This is a contradiction with our intuitive separability requirement for the state at $\gamma \rightarrow J$. 
		
\begin{figure}[h!]
			\centering{\includegraphics[width=0.34\textwidth]{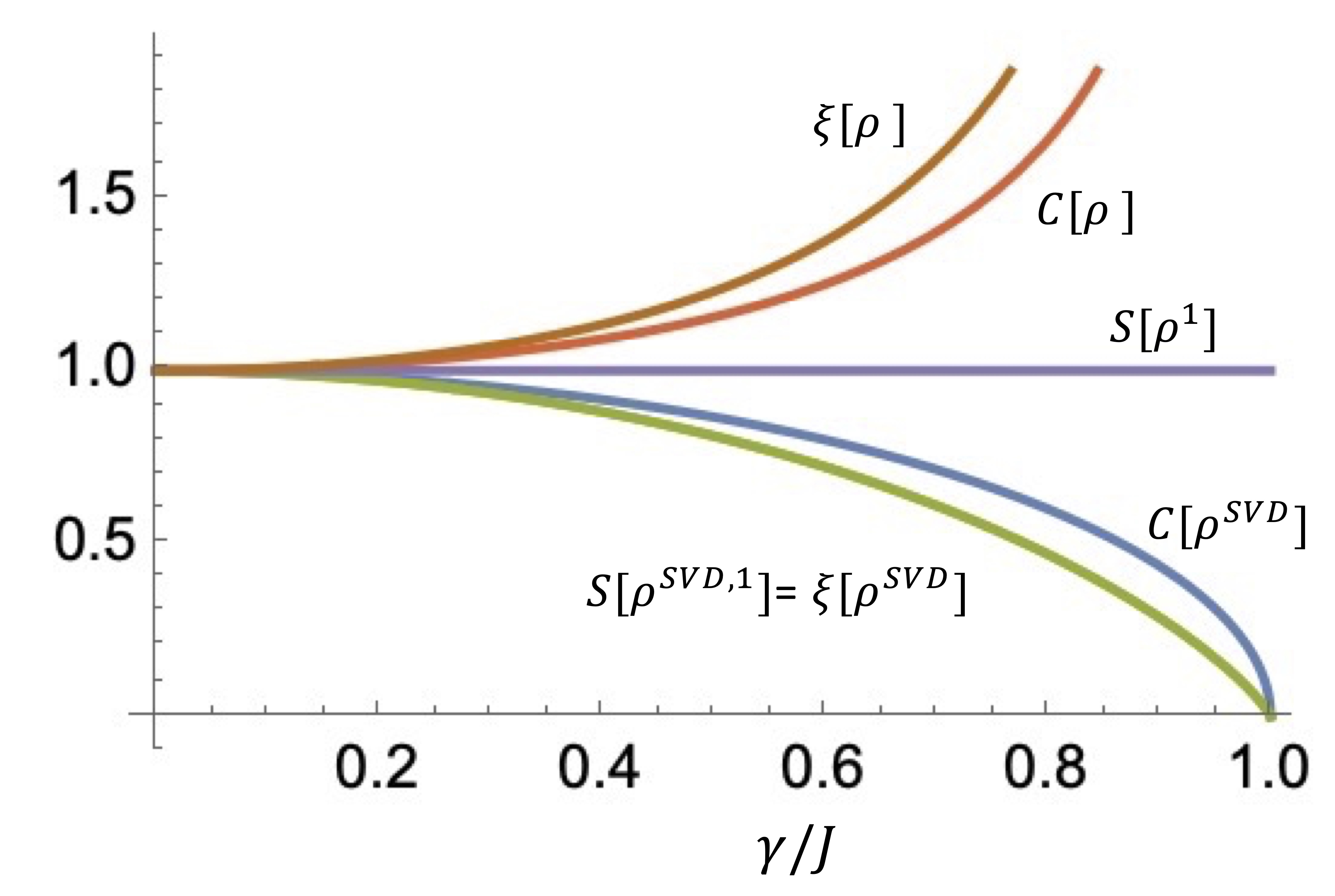}}
			\caption{{Various entanglement measures, such as entropy ($S$), concurrence ($C$), and entanglement of formation ($\xi$), are presented for the bi-orthogonal pure state $\lvert R_0\rangle\langle L_0\rvert$. It is observed that the conventional density matrix $\rho$ in Eq.~\eqref{Eq-B1} yields inconsistent entanglement values, whereas measures derived from the SVD density matrix $\rho^{SVD}$ in Eq.~\eqref{Eq-B3} provide consistent results.} } \label{fig-s2}
		\end{figure}

		Other measure of entanglement \cite{Wooters1998} e.g. the concurrence $C=\frac{1}{\sqrt{1-(\gamma/J)^2}}$ obtained from eigenvalues $\{\frac{J}{\sqrt{J^2-\gamma^2}},0,0,0\}$ of $R=[\rho (\sigma^y\otimes \sigma^y) \rho^* (\sigma^y\otimes \sigma^y)]^{1/2}$, and the corresponding entanglement of formation \mbox{$\xi(C)=-\frac{1+\sqrt{1+C^2}}{2} \log_2 \frac{1+\sqrt{1+C^2}}{2}-\frac{1+\sqrt{1-C^2}}{2} \log_2 \frac{1-\sqrt{1+C^2}}{2}$}, not only inconsistent with the entanglement entropy $S$, but also both exceed the upper bound of physical entanglement for all $\gamma$ and diverges at $\gamma\rightarrow J$ [see Fig.~\ref{fig-s2}].

		Now, we consider $\rho^{SVD}=\sqrt{\rho^{\dag}\rho}/\rm Tr \sqrt{\rho^{\dag}\rho}$. 
		For the state $|R_0\rangle$, we obtain
		\begin{gather}
			\rho^{SVD}= \frac{1}{2J}\left(\begin{array}{cccc}
				0 &  0&0  & 0 \\
				0& J+\gamma &-\sqrt{J^2-\gamma^2}& 0 \\
				0& -\sqrt{J^2-\gamma^2}&J-\gamma &0  \\
				0& 0 & 0 & 0
			\end{array} \right).\label{Eq-B3}
		\end{gather}
		The reduced density matrix for qubit 1 is now 
		\begin{equation}
			\rho^{\rm SVD,1}= {\rm Tr}_2(\rho^{\rm SVD})=\frac{J+\gamma}{2J} \lvert\uparrow\rangle\langle\uparrow\rvert + \frac{J-\gamma}{2J}\lvert\downarrow\rangle\langle\downarrow\rvert,
		\end{equation}
		with entropy
		\begin{equation}
			 S=-\frac{J+\gamma}{2J} \log_2 \frac{J+\gamma}{2J} - \frac{J-\gamma}{2J} \log_2 \frac{J-\gamma}{2J}
			 \end{equation}
			 now depends on $\gamma$ and decreases monotonically to zero as $\gamma\rightarrow J$ (Fig.~\ref{fig-s2}).  The concurrence, obtained in this case \mbox{$C=\sqrt{1-(\gamma/J)^2}$} from the eigenvalues \mbox{$\left\{\frac{\sqrt{J^2-\gamma^2}}{J},0,0,0\right\}$} of $R=[\rho^{SVD} (\sigma^y\otimes \sigma^y) \rho^{SVD*} (\sigma^y\otimes \sigma^y)]^{1/2}$, and  the corresponding entanglement of formation $\xi(C)$ is consistent with $S$ (as per the requirement explained in Ref.~\cite{Wooters1998}). We have, therefore, considered $\rho^{SVD}$ for concurrence computations whenever bi-orthogonal states are involved in the construction of a density matrix.   Note that in systems satisfying usual orthogonality one has  $\rho^{SVD}=\rho$. 
		
		\section{\label{Appendix-B} Analytical calculation of concurrence at low temperature limit $T\rightarrow0$}
		Entanglement properties of the thermal state [given by Eq.~\eqref{eq-rho}] is too cumbersome to obtain analytically. To gain analytical insights on the low temperature \mbox{$(T\lesssim|E_1-E_0|)$} thermal entanglement, we  approximate the thermal state consisting only ground and first excited states i.e 
		\begin{equation}\rho=\frac{e^{-E_0/T}|R_0\rangle\langle L_0| +e^{-E_1/T}|R_1\rangle\langle L_1|}{e^{-E_0/T}+e^{-E_1/T}},
			\end{equation}
			 which is in a matrix form given by
		\begin{gather}
			\rho= \alpha\left(\begin{array}{cccc}
				e^{J\delta/T} &  0&0  & -e^{J\delta/T} \\
				0& e^{\sqrt{J^2-\gamma^2}/T} &-\sqrt{\frac{J-\gamma}{J+\gamma}} e^{\sqrt{J^2-\gamma^2}/T}& 0 \\
				0& -\sqrt{\frac{J+\gamma}{J-\gamma}} e^{\sqrt{J^2-\gamma^2}/T}&e^{\sqrt{J^2-\gamma^2}/T} &0  \\
				-e^{J\delta/T}& 0 & 0 & e^{J\delta/T}
			\end{array} \right),
		\end{gather}
		where $\alpha=\left[{2({e^{J\delta/T}+e^{\sqrt{J^2-\gamma^2}/T}})}\right]^{-1}$.
		In order to compute the concurrence, we obtain 
		\begin{gather}
			\rho^{SVD}=\frac{\lambda}{2} \left(\begin{array}{cccc}
				e^{J\delta/T} &  0&0  & -e^{J\delta/T} \\
				0& \sqrt{\frac{J+\gamma}{J-\gamma}} e^{\sqrt{J^2-\gamma^2}/T} &- e^{\sqrt{J^2-\gamma^2}/T}& 0 \\
				0& - e^{\sqrt{J^2-\gamma^2}/T}& \sqrt{\frac{J-\gamma}{J+\gamma}}e^{\sqrt{J^2-\gamma^2}/T} &0  \\
				-e^{J\delta/T}& 0 & 0 & e^{J\delta/T}
			\end{array} \right),
		\end{gather}
		and the corresponding 
		\begin{gather}
			R= \frac{\lambda}{2} \left(\begin{array}{cccc}
				e^{J\delta/T} &  0&0  & -e^{J\delta/T} \\
				0& e^{\sqrt{J^2-\gamma^2}/T} &- \sqrt{\frac{J+\gamma}{J-\gamma}} e^{\sqrt{J^2-\gamma^2}/T}& 0 \\
				0& - \sqrt{\frac{J-\gamma}{J+\gamma}} e^{\sqrt{J^2-\gamma^2}/T}& e^{\sqrt{J^2-\gamma^2}/T} &0  \\
				-e^{J\delta/T}& 0 & 0 & e^{J\delta/T}
			\end{array} \right),
		\end{gather}
		where $\lambda=\sqrt{J^2-\gamma^2}\left[(\sqrt{J^2-\gamma^2}{e^{J\delta/T}+J e^{\sqrt{J^2-\gamma^2}/T}})\right]^{-1}$. Eigenvalues of $R$ and the corresponding concurrence reduces to  Eq.~\eqref{eq-eigv} and first of Eq.~\eqref{eq-QPT} in the main text. 
		To derive the concurrence at $T=0$, we note that $C$ can also be written as
		\begin{equation}
			{C}= \left\{\begin{array}{c}
				\frac{\sqrt{J^2-\gamma^2}}{J+\sqrt{J^2-\gamma^2}e^{-\left(\sqrt{J^2-\gamma^2}J-\delta\right)/T}}~ |e^{-\left(\sqrt{J^2-\gamma^2}-J\delta\right)/T}-1|, ~ \sqrt{J^2-\gamma^2}-J\delta>0	\\
				\frac{\sqrt{J^2-\gamma^2}}{J e^{-\left(J\delta-\sqrt{J^2-\gamma^2}\right)/T}+\sqrt{J^2-\gamma^2}} ~|1-e^{-\left(J\delta-\sqrt{J^2-\gamma^2}\right)/T}|, ~ J\delta -\sqrt{J^2-\gamma^2}>0.
			\end{array}  \right.
		\end{equation}
Hence, the second of Eq.~\eqref{eq-QPT} in the main text readily follows.  
		
{\section{\label{Appendix-C} General solution of $H$ for arbitrary $\gamma_1$ and $\gamma_2$}
		For generic values of $\gamma_1$ and $\gamma_2$, the asymmetric non-Hermitian Hamiltonian $H$ reduces to 
		\begin{equation}
H=\left(\begin{matrix}
	0&0  &0  & J\delta  \\
	0& 0 & J+\gamma_2&0  \\
	0& J+\gamma_1 & 0 &0  \\
	J\delta&0  &0  &0 
\end{matrix}			\right).
			\end{equation}
The right- and left-eigenvectors are given by
	\begin{align}
	& 	\scalebox{.95} {$ 	|R_{0,3}\rangle: \frac{1}{\sqrt{2}}\left( \sqrt{\frac{J+\gamma_2}{J+\gamma_1}}\lvert\uparrow\downarrow\rangle \mp \lvert\downarrow\uparrow\rangle \right),	|R_{1,2}\rangle: \frac{1}{\sqrt{2}}\left( \lvert\uparrow\uparrow\rangle \mp \lvert\downarrow\downarrow\rangle \right) $} \nonumber\\ 
	&  \scalebox{.95} {$ 	|L_{0,3}\rangle: \frac{1}{\sqrt{2}}\left( \sqrt{\frac{J+\gamma_1}{J+\gamma_2}}\lvert\uparrow\uparrow\rangle \mp \lvert \downarrow\downarrow\rangle \right),  ~ \lvert L_{1,2}\rangle = \lvert R_{1,2}\rangle $},
\end{align}
with corresonding energies
\begin{equation*}\label{eq-spectrum}
	E_{0,1,2,3}=\left\{-\sqrt{(J+\gamma_1)(J+\gamma_2)},-J \delta,J \delta, \sqrt{(J+\gamma_1)(J+\gamma_2)}\right\}.
\end{equation*}
The ground state of the system is determined by the minimum value between \(E_0\) and \(E_1\), which depends on the parameters \(\gamma_{1,2}\) and \(\delta\). When \(E_0 < E_1\), specifically when \mbox{$(J+\gamma_1)(J+\gamma_2) > J^2 \delta^2$}, the ground state corresponds to the non-maximally entangled state \(\lvert R_0\rangle\). Conversely, if \mbox{$E_1 < E_0$}, meaning \mbox{$(J+\gamma_1)(J+\gamma_2) < J^2 \delta^2$}, the ground state is the Bell state $\lvert R_1\rangle$. The concurrence for the Bell state satisfies \mbox{$C(\lvert R_1\rangle\langle L_1\rvert) = 1$}, whereas the concurrence for the non-maximally entangled state \(\lvert R_0\rangle\) is given by Eq.~\eqref{Eq-Concurrence} in the main text which is always less than one. As a result, both the nature of the ground state and its entanglement properties at zero temperature (\(T=0\)) undergo changes at parameter values where \(E_0 = E_1\), specifically when \mbox{$(J+\gamma_1)(J+\gamma_2) = J^2 \delta^2$} is satisfied.
}
		
		\vspace{0.0cm}
\subsection*{Acknowledgments} \noindent The research was supported by the ANRF~Grant  (MTR/2023/000249) and a Seed Grant from IISER~Berhampur, India.\\


	\end{document}